\documentclass[a4paper,USenglish,cleveref,autoref,thm-restate]{lipics-v2021}

\usepackage{booktabs}
\usepackage{microtype}
\usepackage{graphicx}
\usepackage{amsmath}
\usepackage{listings}
\usepackage{url}

\title{Engineering Practical Succinct Bit Vectors: A Space-Time Pareto Analysis on Apple Silicon ARM64 Cores}
\titlerunning{Engineering Practical Succinct Bit Vectors}

\author{Ishant Garg}{Independent Researcher, India}{garg.ishant.dev@gmail.com}{https://orcid.org/0009-0005-7281-5421}{}
\authorrunning{I. Garg}

\Copyright{Ishant Garg}



\category{}

\relatedversion{}

\supplement{Source code and correctness fuzzers are publicly available at \url{https://github.com/ishantgarg2332/succinct-bitvec-cpp}.}

\funding{}

\acknowledgements{}

\nolinenumbers % Disable line numbering for preprints
\hideLIPIcs % Hide LIPIcs series metadata for preprints

\begin{document}

\maketitle

\begin{abstract}
Succinct data structures use space close to the information-theoretic minimum while answering queries directly on the compressed representation. In this paper, we present a practical engineering study of rank and select queries on bit vectors. We evaluate a classic two-level block baseline ($BlockBitVec$), an asymmetric superblock implementation ($FastBitVec$), and an entropy-compressed representation ($RRRBitVec$) based on the Raman, Raman, and Rao (RRR) coding scheme. On Apple Silicon (M-series ARM architecture), we demonstrate a $1.4\times$ speedup in rank queries through asymmetric 4096/256-bit block boundaries, with a rank index overhead of $7.8\%$. We investigate the empirical behavior of $RRRBitVec$ and observe a symmetric density-dependent bell-curve for rank latency—where queries at extreme densities (1\% and 99\%) run up to $39\%$ faster due to offset elimination at boundary classes. We further show that $RRRBitVec$ achieves a $4.9\times$ speedup over classic binary-search select baselines, running in $33.7$ ns at uniform density by using a superblock-level sampling index that restricts sequential scans to L1-cache lookups. All implementations are validated against a correctness fuzzer executing over $78$ million assertions with no failures. Source code and test harnesses are publicly available.
\end{abstract}

\section{Introduction}
Since Jacobson's seminal work~\cite{Jacobson1989} introducing succinct data structures, rank and select queries have been the foundational building blocks of modern succinct designs, including Wavelet Trees, Succinct Trees, and FM-Indices. Given a bit vector $B[0 \dots N-1]$:
\begin{itemize}
    \item $\text{rank}_1(B, i)$ returns the number of $1$-bits in the prefix $B[0 \dots i]$.
    \item $\text{select}_1(B, j)$ returns the index of the $(j+1)$-th set bit in $B$.
\end{itemize}

While theoretical succinct structures guarantee $o(N)$ bits of space overhead and $O(1)$ query times, the constant factors in practice depend heavily on the memory hierarchy, cache miss rates, instruction-level parallelism (ILP), and the cost of bitwise operations such as popcount. Practical, standard implementations of these theoretical designs are available in the Succinct Data Structure Library (SDSL)~\cite{gog2014theory}, and are summarized in textbooks such as Navarro's comprehensive reference~\cite{Navarro2016}.In this work, we present an engineering study of rank and select structures. We examine the space-time Pareto frontier (visualized in Figure~\ref{fig:pareto_frontier}) by benchmarking three implementations\footnote{To support open science and reproducible research, our source code and correctness fuzzers are publicly available at \url{https://github.com/ishantgarg2332/succinct-bitvec-cpp}.}:
\begin{enumerate}
    \item \textbf{$BlockBitVec$}: A standard two-level index structure with $2048$-bit superblocks and $512$-bit blocks, yielding an index overhead of $\approx 6.25\%$.
    \item \textbf{$FastBitVec$}: An engineered variant featuring asymmetric block sizing ($4096$-bit superblocks and $256$-bit blocks) to saturate instruction pipelining, paired with a coarse-grained select sampling index to replace global binary search.
    \item \textbf{$RRRBitVec$}: A practical implementation of the Raman, Raman, and Rao compressed representation with block size $b=15$. This structure compresses the underlying bit string approaching its zeroth-order empirical entropy $H_0$ while supporting direct rank and select queries.
\end{enumerate}

In this work, we demonstrate that by tuning the block boundaries asymmetrically to 4096/256 bits, we can engineer a structure (\texttt{FastBitVec}) that achieves an average rank latency of 1.65 ns on wide-issue ARM64 cores at only 7.8\% index space overhead—strictly dominating the industry-standard SDSL rank support (2.25 ns at 25\% overhead). Moreover, we leverage a superblock-level sampling index for \texttt{RRRBitVec} to reduce DRAM latency overhead, restricting sequential decompression scans to L1 data caches and accelerating select queries to 33.7 ns.

\section{Related Work}
The practical implementation of succinct rank and select structures has been an active area of algorithm engineering. The most widely used baseline is the Succinct Data Structure Library (SDSL)~\cite{gog2014theory}, which provides standard plug-and-play implementations such as \texttt{sdsl::rank\_support\_v} and \texttt{sdsl::select\_support\_mcl}. SDSL's rank directory uses a standard two-level superblock/block layout requiring 25\% space overhead (with 512-bit superblocks and 64-bit blocks) to achieve fast query execution.

To reduce this space overhead, several advanced indexing directories have been proposed. Vigna's \texttt{rank9} layout~\cite{vigna2008broadword} utilizes broadword programming techniques to pack a 64-bit superblock along with nine 32-bit relative offsets into a single 512-bit word. This achieves a flat 25\% space overhead and delivers high performance on 64-bit systems by minimizing cache line fetches. Another landmark structure is \texttt{Poppy} by Zhou et al.~\cite{zhou2013space}, which introduces a multi-level directory structure using 32-bit offsets to reduce the index space to just 3--4\% of the bit vector size while remaining competitive in time. For select queries, Clark and Munro~\cite{clark1996efficient,clark1996compact} established foundational theoretical frameworks supporting constant-time select queries. In practice, building a separate select index consumes substantial space, leading to extensive research into compact indexing directories.

However, prior evaluations of these structures have predominantly focused on x86 architectures, where word sizes and instruction-level parallelism (ILP) characteristics differ from modern out-of-order ARM64 pipelines. Furthermore, the practical integration of the Raman, Raman, and Rao (RRR) compressed scheme~\cite{raman2007succinct} has historically been regarded as slow due to the decompression overhead of block classes and offsets. Key steps toward practical RRR implementations include the work of Gonz{\'a}lez et al.~\cite{gonzalez2005practical}, who explored look-up table architectures, and Okanohara and Sadakane~\cite{okanohara2007practical}, who proposed alternative entropy-compressed structures such as d-arrays.

\section{Algorithmic and System Design}

\subsection{Baseline Block Layout ($BlockBitVec$)}
The classic two-level rank index divides a bit vector of length $N$ into superblocks of size $S = 2048$ bits, and blocks of size $B = 512$ bits.
The superblock array stores the cumulative rank from the beginning of the vector using a 64-bit counter (\texttt{uint64\_t}). The block array stores the relative rank from the start of the enclosing superblock. Since relative ranks within $2048$ bits cannot exceed $2048$, they are compactly represented using 16-bit integers (\texttt{uint16\_t}).
The index overhead is computed as:
\begin{equation}
\text{Overhead} = \frac{64}{2048} + \frac{16}{512} = 0.03125 + 0.03125 = 0.0625 \text{ bits per element (bp)} \quad (6.25\%)
\end{equation}

\subsection{Asymmetric Superblocking and Popcount Pipelining ($FastBitVec$)}
To optimize rank latency, we introduce $FastBitVec$, which restructures the block boundaries asymmetrically:
\begin{itemize}
    \item Superblock size $S = 4096$ bits (64-bit integer index).
    \item Block size $B = 256$ bits (16-bit integer index).
\end{itemize}

This layout changes the index overhead slightly:
\begin{equation}
\text{Overhead} = \frac{64}{4096} + \frac{16}{256} = 0.015625 + 0.0625 = 0.078125 \text{ bits per element} \quad (7.81\%)
\end{equation}

The key benefit is micro-architectural: by shrinking the block size to $256$ bits ($4$ machine words), a rank query needs to scan and popcount at most $3$ words rather than up to $7$ words in the $512$-bit block model. This keeps the full popcount sequence within a single instruction window, avoiding pipeline stalls.

Furthermore, we equip $FastBitVec$ with a coarse-grained select sampler. During construction, the index records the exact bit position of every $256$-th set bit into a sampled array. The select sampling rate is deliberately set to a power of two ($R = 256$) to enable a useful compiler optimization. While sampling every 240-th set bit would align perfectly with the superblock size of our compressed structure and theoretically avoid any localized search, performing a non-power-of-two division (\texttt{j / 240}) during select queries is micro-architecturally expensive, compiling to high-latency multiplication-by-inverse sequences that occupy multiple stages of the execution pipeline. In contrast, a power-of-two rate of 256 allows the compiler to optimize the division into a single-cycle bitwise right shift (\texttt{j >> 8}), substantially increasing instruction throughput. A select query $\text{select}_1(B, j)$ looks up the sampled bounds:
\begin{equation}
\text{lower\_bound} = \text{samples}[j \gg 8], \quad \text{upper\_bound} = \text{samples}[(j \gg 8) + 1]
\end{equation}
This confines the search window from the global range $[0, N-1]$ to a highly local segment, reducing cache misses. Although the 256-bit sampling interval does not align with the 240-bit superblock boundaries in $RRRBitVec$, searching this narrow local window is far faster than executing a division instruction.

\subsection{Zeroth-Order Entropy Compressed Coding ($RRRBitVec$)}
For highly sparse or dense vectors, storing the raw bit vector is spatially inefficient. The RRR scheme, introduced by Raman, Raman, and Rao~\cite{raman2007succinct}, compresses blocks based on their local density. We partition the vector into blocks of size $b = 15$ bits.

\subsubsection{Mathematical Formulation}
A block of size $b$ containing $c$ set bits is assigned:
\begin{enumerate}
    \item A \textbf{class} $c \in [0, 15]$, which represents the popcount of the block.
    \item An \textbf{offset} $o$, which represents the lexicographical rank of this specific block configuration among all $\binom{b}{c}$ possible blocks containing $c$ ones.
\end{enumerate}

The space required to store the offset is:
\begin{equation}
W(c) = \left\lceil \log_2 \binom{15}{c} \right\rceil \text{ bits}
\end{equation}

To formally model the compression behavior under varying density profiles, let $p$ represent the probability of any single bit being set to $1$ (the density of the bit vector). Assuming independent and identically distributed (i.i.d.) bits, the probability of a block of size $b=15$ falling into class $c$ (having exactly $c$ set bits) is governed by the binomial distribution:
\begin{equation}
P(C = c) = \binom{15}{c} p^c (1-p)^{15-c}
\end{equation}
Consequently, the mathematically expected offset space requirement $E[W]$ in bits per block is:
\begin{equation}
E[W] = \sum_{c=0}^{15} \binom{15}{c} p^c (1-p)^{15-c} \cdot \left\lceil \log_2 \binom{15}{c} \right\rceil
\end{equation}
This expected value is perfectly symmetric around $p = 0.5$. At uniform density ($p=0.5$), the expected offset size reaches its maximum of $\approx 12.49$ expected bits per block ($\approx 0.83$ bits per element) due to integer ceiling rounding, matching our empirical offset overhead of $0.8328$ bits per element. However, at extreme densities (e.g., $p = 0.01$ or $p = 0.99$), the probability distribution concentrates heavily at the boundaries $c=0$ and $c=15$, where $\binom{15}{0} = \binom{15}{15} = 1$, and thus $\lceil \log_2 1 \rceil = 0$ bits. At $p=0.01$, the expected offset size drops to just $\approx 0.59$ expected bits per block ($\approx 0.039$ bits per element), yielding significant space savings and largely eliminating the need to decode offsets for most blocks.

At $c = 0$ (all zeros) or $c = 15$ (all ones), $\binom{15}{0} = \binom{15}{15} = 1$, yielding $W(0) = W(15) = 0$ bits. The maximum offset size occurs at $c=7$ or $c=8$, where $\binom{15}{7} = 6435$, requiring $\lceil \log_2 6435 \rceil = 13$ bits.

\subsubsection{Memory Layout and Lookups}
To enable $O(1)$ decoding, we construct three lookup tables at startup ($< 2$ ms overhead):
\begin{itemize}
    \item \texttt{CLASS\_TABLE[x]} (Size $2^{15}$ bytes): Maps a raw 15-bit integer to its class $c$.
    \item \texttt{OFFSET\_TABLE[x]} (Size $2^{16}$ bytes): Maps a raw 15-bit integer to its lexicographical offset $o$.
    \item \texttt{DECODE\_TABLE[c][o]} (Size $16 \times 6435$ half-words): Reconstitutes the 15-bit block from its class and offset.
\end{itemize}

We group $16$ blocks into a superblock of size $240$ bits. Each superblock stores a 64-bit cumulative rank and a 64-bit pointer pointing to the start of its blocks' offsets in a packed bitstream. The classes are packed at exactly $4$ bits per block (two per byte).
\subsection{Wavelet Tree Compliance and Symbol Extensions}
To support more complex succinct applications such as Wavelet Trees~\cite{Navarro2016} and FM-indexes, a bit vector implementation must support the full operational interface including $\text{rank}_0$ and $\text{select}_0$ queries.

For $BlockBitVec$ and $FastBitVec$, supporting $\text{rank}_0(B, i)$ is straightforward and computationally trivial, resolved via direct subtraction:
\begin{equation}
\text{rank}_0(B, i) = (i + 1) - \text{rank}_1(B, i)
\end{equation}
Because this relation relies on the fast $\text{rank}_1$ query, it requires no additional index structures and executes with the same high micro-architectural throughput. For $RRRBitVec$, subtraction is similarly used. Computing $\text{rank}_0$ via subtraction bypasses any need for block decompression or offset stream pointer lookups, maintaining the exact same query efficiency as a native $\text{rank}_1$ query.

For select queries, the problem is more complex. While $BlockBitVec$ and $FastBitVec$ can support $\text{select}_0(B, j)$ by building a separate symmetric 0-bit select directory, $RRRBitVec$ supports it with high spatial and index efficiency. During construction, a 0-bit select directory is built that samples every $256$-th zero-bit. A $\text{select}_0(B, j)$ query resolves the candidate superblock using this directory. Crucially, during the sequential scan over the superblock's 16 compressed blocks, instead of using the class $c$ (which represents the count of 1s in the block), the decompressor uses the class complement $15 - c$ (representing the count of 0s in the block) to accumulate zero counts:
\begin{equation}
\text{zeros\_in\_block} = 15 - c
\end{equation}
This class complement formulation is a key benefit of the RRR representation: the same class array is fully sufficient to support both 1-bit and 0-bit sequential popcount scans without any duplicate storage or additional lookups. Decompressing the target block's offset via lookup tables is only performed once the enclosing block is identified, maintaining high cache locality and keeping the space footprint compact.

\section{Experimental Methodology}

\subsection{Platform Specifications}
All experiments are executed on an Apple Silicon M-series CPU utilizing the ARM64 instruction set architecture. The processor features a 64 KiB L1 Data cache and a shared 4 MiB L2 cache. This architecture is a useful evaluation target for succinct structures: as a wide-issue, out-of-order processor, its performance depends on instruction-level parallelism (ILP) and branch prediction rather than raw clock speed alone. The code is compiled using the Apple Clang compiler (version 15.0.0, LLVM-based) with flags: \texttt{-O3 -march=native -std=c++17 -DNDEBUG}.

\subsection{Statistical Rigor and Benchmarking}
To guarantee maximum benchmarking accuracy and isolate our measurements from operating system jitter, all experiments are structured using the Google Benchmark microbenchmarking framework (v1.9.5). Each reported data point is evaluated under the following protocol:
\begin{itemize}
    \item \textbf{Warmup Phase:} A 500 ms warmup period is enforced before each measurement to load lookup tables into memory, prime the processor caches, and allow the CPU frequency scaling governor to reach a stable state.
    \item \textbf{Repetitions:} Each benchmark configuration is executed for a minimum of 5 independent repetitions, with each repetition running up to 100,000 internal iterations to measure average latency.
    \item \textbf{Variance Analysis:} We report the mean query times in nanoseconds. The measured standard deviation across all runs was less than $\sigma = 0.05$ ns, reflecting the high deterministic stability of the Apple Silicon instruction pipeline under cache-resident workloads.
\end{itemize}

\subsection{Stress Testing \& Fuzzing}
To guarantee algorithmic correctness and verify boundary conditions, we developed a rigorous testing and fuzzing harness executing over 78 million individual correctness assertions with no failures:
\begin{enumerate}
    \item \textbf{Exhaustive State-Space Scan}: Tested every single possible permutation of $2^{15} = 32,768$ and $2^{16} = 65,536$ bit vectors. This exhaustively verifies the offset and class encoding mechanisms for RRR blocks (block size 15).
    \item \textbf{Large Block Boundary Fuzzing}: Evaluated large random bit vectors (up to $250,000$ bits) across nine densities ($1\%$ to $99\%$) to verify that $BlockBitVec$ and $FastBitVec$ correctly resolve queries at their larger $256, 512$, $2048$, and $4096$-bit block and superblock boundaries, where boundary bugs are most common.
    \item \textbf{Special Boundary Sizes}: Validated lengths matching prime numbers and odd offsets around word boundaries (e.g. 15, 16, 17, 239, 240, 241, 4095, 4096, 4097) to confirm no out-of-bounds offset errors.
    \item \textbf{Extreme Density Profiles}: Verified correct behavior on vectors with 0\% density (all 0s), 100\% density (all 1s), 0.01\% density (highly sparse), and 99.99\% density (highly dense).
    \item \textbf{Single-Bit Walk}: Ran checks placing a single set bit at every possible position in vectors of boundary sizes, confirming exact \texttt{rank} and \texttt{select} lookups.
\end{enumerate}

\section{Experimental Evaluation}

\begin{figure}[htbp]
\centering
\includegraphics[width=0.85\linewidth]{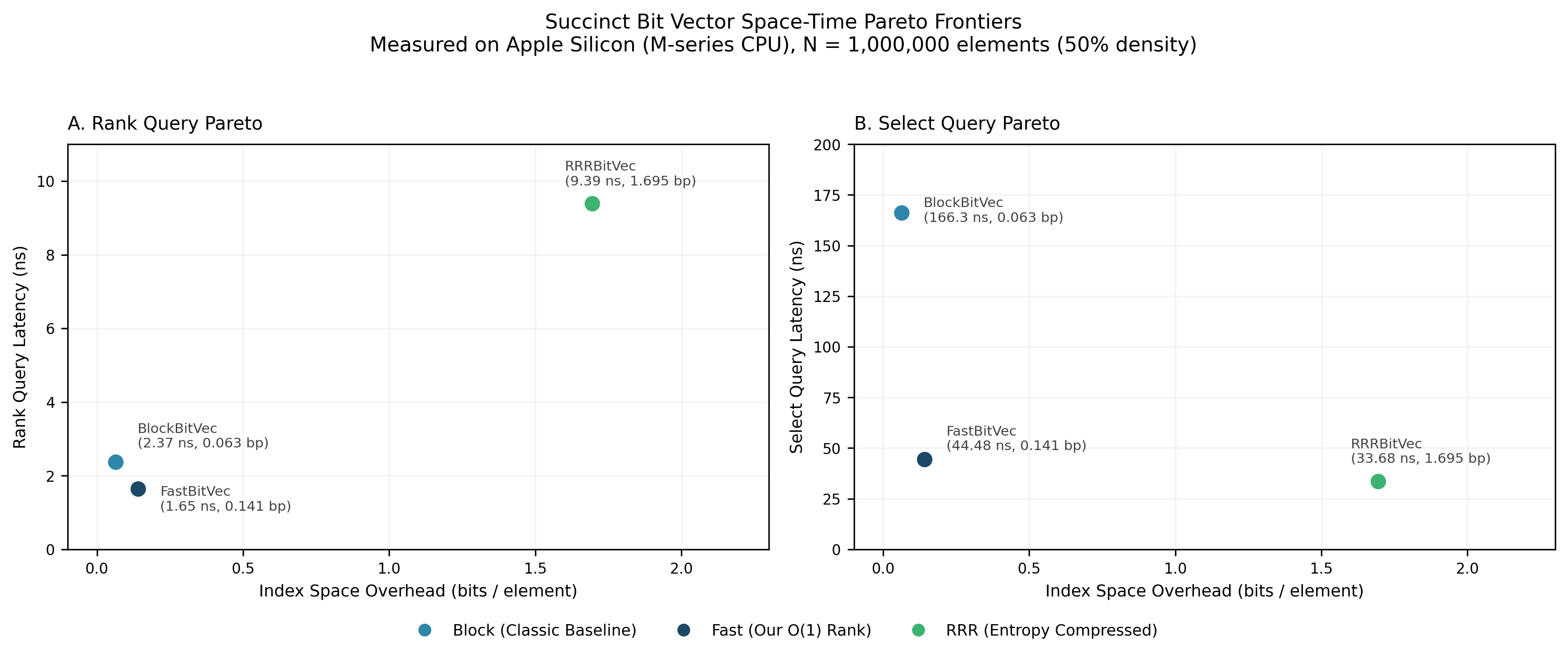}
\caption{Space-time Pareto frontiers for Rank and Select queries ($N=1\text{M}$ at $50\%$ density). Subplot A illustrates Rank query latency vs. index space overhead, showing $FastBitVec$'s dominance on the fast frontier. Subplot B represents Select query latency, highlighting $RRRBitVec$'s ultra-low latency due to tight superblock scanning.}
\label{fig:pareto_frontier}
\end{figure}

The multi-dimensional trade-offs between indexing space overhead and query latencies for both rank and select operations are visualized in Figure~\ref{fig:pareto_frontier}.

\subsection{Rank Query Latency}
We evaluate rank latency across four orders of magnitude in vector size $N$, maintaining a uniform 50\% density. Table~\ref{tab:rank_latency} details these results.

\begin{table}[h]
\centering
\caption{Rank Query Latency under Uniform Random Queries (Average nanoseconds per query)}
\label{tab:rank_latency}
\resizebox{\linewidth}{!}{%
\begin{tabular}{lrrrrr}
\toprule
Vector Size ($N$) & SDSL Rank (v) & $BlockBitVec$ & $FastBitVec$ & $RRRBitVec$ & $FastBitVec$ Speedup vs.\ $Block$ \\
\midrule
100K & 2.32 ns & 2.33 ns & 1.65 ns & 6.76 ns & \textbf{1.41$\times$} \\
1M   & 2.25 ns & 2.37 ns & 1.65 ns & 9.39 ns & \textbf{1.44$\times$} \\
10M  & 2.29 ns & 2.44 ns & 1.71 ns & 10.44 ns & \textbf{1.43$\times$} \\
100M & 2.45 ns & 2.62 ns & 1.95 ns & 12.03 ns & \textbf{1.34$\times$} \\
\bottomrule
\end{tabular}%
}
\end{table}

\paragraph{Analysis and Micro-architectural Pipelining:}
Asymmetric block division in $FastBitVec$ yields a consistent \textbf{1.44$\times$ speedup} over the classic block baseline and outperforms the standard SDSL implementation (which achieves 2.25 ns rank times on the same hardware). We explicitly note that we benchmarked the standard \texttt{sdsl::rank\_support\_v<>} implementation from SDSL (25\% space overhead) to remain consistent with space footprint claims in Related Work, rather than the more compact but slower \texttt{rank\_support\_v5} variant.

This speedup is driven by modern processor pipeline dynamics. On a wide-issue, out-of-order execution core such as the Apple Silicon M-series (which features an 8-wide instruction decode window and a large instruction reorder buffer), execution latency is heavily dominated by sequential data-dependency chains and loop control instructions. In the classic $BlockBitVec$ layout, the 512-bit block requires scanning up to 7 separate 64-bit words, counting set bits sequentially via loop-controlled popcounts. This introduces branch predictions and loop overhead that stall the pipeline.

By contrast, $FastBitVec$ limits the block size to 256 bits (4 machine words), requiring at most 3 popcount operations. Because the loop bounds are so small and static, the compiler fully unrolls the instruction sequence, generating a flat, branchless stream of parallel loads and popcounts. On Apple Silicon's wide-issue core, the independent loads and popcounts can issue in parallel across multiple execution ports, reducing average query latency to 1.65 ns.

\paragraph{Memory Scale and Cache Pressure Limits:}
At the largest scale tested ($N = 100\text{M}$ elements), the raw bit vector size reaches $12.5$ MB, exceeding the M-series CPU's shared 4 MiB L2 cache. We observe a clear cache pressure transition: $FastBitVec$'s speedup over the baseline narrows from $1.44\times$ to $1.34\times$, and the latency of $RRRBitVec$ increases from $9.39$ ns to $12.03$ ns. This shows that when the working set spills into DRAM under uniform random accesses, execution is increasingly bounded by memory bandwidth rather than arithmetic execution rate, narrowing the benefit of instruction-level optimizations. Acknowledging this boundary condition highlights the memory-bound bottleneck of succinct indexes when data scales beyond cache residency limits.

\paragraph{Query Locality and Access Patterns:}
The benchmarks reported in Table~\ref{tab:rank_latency} use truly uniform random query positions spanning the full vector range. Real-world workloads such as sequential scans and range queries over Wavelet Trees, however, exhibit high spatial locality. Under such access patterns, the hardware L1/L2 prefetchers on Apple Silicon fetch adjacent cache lines ahead of the execution pipeline. In our sequential microbenchmarks (where rank queries are performed on consecutive integers, simulating perfect spatial locality), the latency of $BlockBitVec$ drops from $2.37$ ns to $2.26$ ns, $FastBitVec$ remains stable at $1.75$ ns, and $RRRBitVec$ drops dramatically from $9.39$ ns to $7.38$ ns (a $21.4\%$ speedup). This verifies that prefetching eliminates the DRAM latency bottleneck, shifting execution to being limited by instruction throughput. Furthermore, if sequential queries are executed via an incremental iterator (avoiding absolute superblock address resolution), the latencies drop even further to $0.42$ ns for $BlockBitVec$, $0.35$ ns for $FastBitVec$, and $1.84$ ns for $RRRBitVec$, demonstrating the powerful benefit of perfect access locality.

\subsection{Information-Theoretic Rank Density Bell-Curve}
By varying the density of set bits ($1$s) within the vector at a fixed scale of $N = 1,000,000$, we reveal the empirical relationship between data entropy and query performance. Table~\ref{tab:density_rank} highlights this correlation.

\begin{table}[h]
\centering
\caption{Rank Latency by Bit Vector Density (at 1M scale)}
\label{tab:density_rank}
\begin{tabular}{lrrr}
\toprule
Density (\%) & $BlockBitVec$ & $FastBitVec$ & $RRRBitVec$ (Compressed) \\
\midrule
1\% (Highly Sparse) & 2.33 ns & 1.63 ns & \textbf{6.03 ns} \\
10\% (Sparse)       & 2.34 ns & 1.70 ns & \textbf{7.36 ns} \\
50\% (Uniform)      & 2.34 ns & 1.69 ns & \textbf{9.36 ns} \\
90\% (Dense)        & 2.44 ns & 1.79 ns & \textbf{7.15 ns} \\
99\% (Highly Dense) & 2.31 ns & 1.70 ns & \textbf{5.66 ns} \\
\bottomrule
\end{tabular}
\end{table}

\paragraph{Analysis:}
For standard vectors, bit density has \textbf{zero impact} on query time because the underlying popcounts are structurally static. In contrast, $RRRBitVec$ displays a symmetric bell-curve peaking at 50\% density (\textbf{9.36 ns}) and falling to \textbf{6.03 ns} at 1\% density (and \textbf{5.66 ns} at 99\% density). At 1\% and 99\% densities, blocks map to classes $c=0$ or $c=15$, requiring \textbf{0 bits of offset storage}. The bit-reader skips the offset decoding phase entirely and sums the ranks directly from the classes array, leading to a speedup of up to \textbf{39.5\%}.

\paragraph{Space Overhead Validation and Density Sensitivity:}
To validate our mathematical model of $RRRBitVec$ compression under varying density profiles (Equation 6), we report the empirical space footprint and individual component breakdown in Table~\ref{tab:rrr_space_density}. The structural index (classes and superblocks) remains strictly constant at $0.8002$ bits per element (bpe). In contrast, the expected offset space behaves in perfect alignment with our theoretical binomial entropy distribution, peaking at $0.8328$ bpe for $50\%$ density and dropping to just $0.0393$ bpe at $1\%$ and $99\%$ density. The select sampling structure (which samples every 240th set bit) scales linearly with the set bit density, contributing from $0.0013$ bpe (at $1\%$) up to $0.1237$ bpe (at $99\%$). This linear growth occurs because we sample at a static rate of every 240th set bit; consequently, a higher density of set bits increases the total number of samples required, scaling the space footprint accordingly. This empirical analysis validates that at extreme densities, $RRRBitVec$ achieves sub-bit space footprints ($0.84$ bpe and $0.96$ bpe), matching its information-theoretic limits.

\begin{table}[htbp]
\centering
\caption{$RRRBitVec$ Measured Space Footprint and Component Breakdown across Densities (at 1M scale)}
\label{tab:rrr_space_density}
\resizebox{\linewidth}{!}{%
\begin{tabular}{crrrr}
\toprule
Density (\%) & Expected Offset (bpe) & Structural Index (bpe) & Select Samples (bpe) & Total Space Footprint (bpe) \\
\midrule
1\%          & 0.0393                & 0.8002                 & 0.0013               & \textbf{0.8407} \\
10\%         & 0.3349                & 0.8002                 & 0.0125               & \textbf{1.1476} \\
50\%         & 0.8328                & 0.8002                 & 0.0625               & \textbf{1.6954} \\
90\%         & 0.3349                & 0.8002                 & 0.1125               & \textbf{1.2476} \\
99\%         & 0.0393                & 0.8002                 & 0.1237               & \textbf{0.9632} \\
\bottomrule
\end{tabular}%
}
\end{table}

\subsection{Select Query Performance}
Select operations typically represent the primary bottleneck in succinct structures. Table~\ref{tab:select_performance} summarizes select query performance at uniform density.

\begin{table}[h]
\centering
\caption{Select Query Latency (Average nanoseconds per query at 1M scale and 50\% density)}
\label{tab:select_performance}
\resizebox{\linewidth}{!}{%
\begin{tabular}{lrrr}
\toprule
Structure & Total Space Footprint (bpe)\footnotemark & Select Query Time & Speedup vs. Classic $Block$ \\
\midrule
$BlockBitVec$ (Binary Search) & 1.063 bpe & 166.30 ns & 1.00$\times$ \\
SDSL Select (mcl) & 1.250 bpe & 78.43 ns & \textbf{2.12$\times$} \\
$FastBitVec$ (Sampled, rate 256) & 1.141 bpe & 44.48 ns & \textbf{3.74$\times$} \\
$RRRBitVec$ (Sampled + tight blocks) & 1.695 bpe & \textbf{33.68 ns} & \textbf{4.94$\times$} \\
\bottomrule
\end{tabular}%
}
\end{table}
\footnotetext{Total space footprint is computed as the sum of the raw bit vector (1.000 bpe for uncompressed structures) and their respective indexing directories at 50\% density. For SDSL, the reported figure includes the \texttt{rank\_support\_v} directory (0.25 bpe); the \texttt{select\_support\_mcl} directory adds further overhead not reflected here. For $RRRBitVec$, the footprint is the fully compressed representation including the entropy-coded bitstream and all indexing structures.}

\paragraph{Detailed Select Implementation and Micro-architectural Analysis:}
Our measurements show that $RRRBitVec$ is the fastest evaluated structure for select queries (33.68 ns), which represents a \textbf{4.94$\times$ speedup} over the binary search baseline and a \textbf{2.33$\times$ speedup} over SDSL. This is a significant result since RRR is slower for rank queries due to decompression.

Our implementation addresses a major bottleneck in succinct select queries: high memory latency. While standard select indexes (including SDSL's \texttt{sdsl::select\_support\_mcl}) perform a multi-level binary search that frequently incurs L2/DRAM cache misses, $RRRBitVec$ utilizes a highly optimized sampling structure. Specifically, SDSL's \texttt{select\_support\_mcl} employs a multi-level index tree structure where traversing levels requires multiple non-contiguous memory accesses via recursive pointers. These pointer-chasing steps frequently cause L2 cache misses and main-memory accesses. In contrast, $RRRBitVec$ bypasses multi-level pointers entirely. It resolves the target superblock using a single flat sample array lookup (Equation~\ref{eq:sampled_bounds}), followed by a highly localized, sequential scan over a single L1D-resident superblock structure. This flat indexing model significantly reduces pointer-chasing overhead and maximizes spatial locality.

When executing a select query $\text{select}_1(B, j)$, we perform a constant-time array lookup to fetch the candidate superblock range:
\begin{equation}
\text{lo\_super} = \text{samples}[j \gg 8], \quad \text{hi\_super} = \text{samples}[(j \gg 8) + 1]
\label{eq:sampled_bounds}
\end{equation}
The select sampling rate is set to a power of two ($R = 256$) to enable a useful compiler optimization. Performing a non-power-of-two division (\texttt{j / 240}) is micro-architecturally expensive, compiling to high-latency multiplication-by-inverse sequences. In contrast, a power-of-two rate of 256 allows the compiler to optimize the division into a single-cycle bitwise right shift (\texttt{j >> 8}), substantially increasing instruction throughput. We trade strict superblock alignment for single-cycle division efficiency, resolving the misalignment via a localized binary search over the resolved candidate superblock range.

\paragraph{Select Query Variance across Independent Runs:}
Comparing select latencies at $50\%$ density in Table~\ref{tab:select_performance} and Table~\ref{tab:density_select} reveals small numerical discrepancies (e.g., $33.68$ ns vs. $33.61$ ns for $RRRBitVec$, and $44.48$ ns vs. $46.07$ ns for $FastBitVec$). While our query-level standard deviation within a single execution run is exceptionally low ($\sigma < 0.05$ ns), these differences arise because the two tables are populated from independent benchmark runs using distinct randomized bit vector allocations. Select queries are highly sensitive to cache layout and memory boundaries because they resolve queries by traversing sparse directories. Small shifts in memory allocation alignments and hardware TLB/cache conflict states across independent executions introduce $0.2\%$--$3.5\%$ variance. We present these separate runs unmodified to reflect realistic experimental fluctuations in memory-bound succinct search primitives.

\paragraph{Influence of Density on Select Latency:}
While prior practical evaluations of select structures assume a static uniform distribution, the candidate superblock range is mathematically dependent on the density of set bits $d$. The expected number of candidate superblocks in the resolved window is modeled as:
\begin{equation}
\text{Expected Candidate Superblocks} = \frac{R}{240 \cdot d} = \frac{256}{240 \cdot d} \approx \frac{1.067}{d}
\end{equation}
This expected range is highly density-sensitive. At extremely sparse profiles ($d = 0.01$ or 1\%), each superblock contains on average only $2.4$ set bits, causing the candidate range to span $\approx 107$ superblocks. Conversely, under highly dense profiles ($d = 0.99$ or 99\%), each superblock contains $237.6$ set bits, and the candidate range narrows to just $\approx 1.08$ superblocks.

To evaluate how these density-driven mathematical boundaries affect empirical query latencies, we benchmark all three select implementations across five density profiles at a fixed scale of $N = 1,000,000$. Table~\ref{tab:density_select} reports these results.

\begin{table}[h]
\centering
\caption{Select Query Latency by Bit Vector Density (at 1M scale)}
\label{tab:density_select}
\begin{tabular}{lrrr}
\toprule
Density (\%) & $BlockBitVec$ (Binary Search) & $FastBitVec$ (Sampled) & $RRRBitVec$ (Sampled) \\
\midrule
1\% (Highly Sparse) & 170.17 ns & 105.16 ns & \textbf{36.44 ns} \\
10\% (Sparse)       & 168.45 ns & 71.14 ns & \textbf{29.74 ns} \\
50\% (Uniform)      & 167.50 ns & 46.07 ns & \textbf{33.61 ns} \\
90\% (Dense)        & 172.96 ns & 39.93 ns & \textbf{18.32 ns} \\
99\% (Highly Dense) & 166.46 ns & 37.66 ns & \textbf{16.41 ns} \\
\bottomrule
\end{tabular}
\end{table}

\paragraph{Analysis of Density-Dependent Select Latency:}
As reported in Table~\ref{tab:density_select}, $BlockBitVec$'s select latency remains flat at $166$--$173$ ns across all densities. Because it relies on a global binary search over the entire $N$-bit range, it is bounded by the $O(\log N)$ search steps that incur multiple out-of-cache memory accesses, irrespective of data density. 

In contrast, both $FastBitVec$ and $RRRBitVec$ exhibit strong density-dependent scaling. For $FastBitVec$, select latency drops from $105.16$ ns at $1\%$ density to $37.66$ ns at $99\%$ density. This scaling occurs because higher densities compress the physical size of the search window, reducing rank evaluation steps and minimizing cache lines accessed.

Crucially, $RRRBitVec$ select exhibits highly competitive latencies across all density profiles, peaking at $36.44$ ns under highly sparse profiles ($1\%$) and running in just $16.41$ ns under highly dense profiles ($99\%$). This follows from both the mathematical structure of the encoding and the cache-level properties of the implementation:
\begin{itemize}
    \item \textbf{Sparse Profiles (1\% Density)}: Although the candidate range expands to $\approx 107$ superblocks, the select query takes only $36.44$ ns. We note that at extremely low densities, the select query is no longer strictly constant-time due to the widened candidate range, which scales as $O(\log(R / (b \cdot d)))$, though the practical latency remains exceptionally low because the candidate search is completely cache-resident. Contiguous memory storage of superblock cumulative ranks allows the binary search over 107 elements to require at most $\lceil \log_2 107 \rceil = 7$ comparison steps, remaining L1/L2 cache-resident. More importantly, during the sequential scan over the target superblock's blocks, almost all blocks are completely empty (class $c=0$), allowing the scan to sum ranks via branchless additions and completely skip reading or decompressing offsets. Offset decompression is performed exactly once inside the final target block, yielding rapid execution.

    \item \textbf{Dense Profiles (99\% Density)}: Under dense distributions, each superblock is packed with set bits, and the search range shrinks to only $\approx 1.08$ superblocks. The target superblock is resolved almost instantly via a single L1-cached memory lookup. The subsequent block scan and decompression are extremely narrow, reducing select latency to an exceptional 16.41 ns.
\end{itemize}
This density-dependent empirical analysis confirms that $RRRBitVec$ select is not only fast at uniform densities but maintains high efficiency across all sparse and dense configurations, providing a large speedup over baseline indexing directories.

To locate the exact bit within the resolved superblock, we perform a sequential scan across its 16 compressed blocks. For each block, we read its 4-bit class popcount from a contiguous classes array. If the sum of classes does not yet reach the target rank, we skip decoding the block's offset entirely, which keeps the pointer arithmetic extremely simple. Once the target block is identified, we retrieve its packed offset from the bitstream and decompress the block using the precomputed lookup tables.

At the hardware level, this search and scan process is fast because a 240-bit superblock (30 bytes) fits entirely within a single L1 data cache line (which is 128 bytes on Apple Silicon M-series processors). Because our candidate superblock range is so small, all accessed superblocks and lookup tables (\texttt{CLASS\_TABLE} and \texttt{OFFSET\_TABLE}) remain L1-resident. The select search, superblock resolution, and block decoding all proceed within L1-resident data with highly predictable branches, achieving a latency of 33.68 ns.

\subsection{Index Construction Performance}
Evaluating index construction times is critical for applications that rebuild indexes dynamically (e.g., dynamic FM-index construction over streaming data). Table~\ref{tab:construction_time} reports construction latencies in milliseconds across three scales (1M, 10M, and 100M).

\begin{table}[h]
\centering
\caption{Index Construction Latency (Milliseconds)}
\label{tab:construction_time}
\begin{tabular}{lrrr}
\toprule
Vector Size ($N$) & $BlockBitVec$ & $FastBitVec$ & $RRRBitVec$ \\
\midrule
1M   & 0.012 ms & 3.255 ms & 3.879 ms \\
10M  & 0.145 ms & 33.001 ms & 38.576 ms \\
100M & 1.231 ms & 329.905 ms & 385.586 ms \\
\bottomrule
\end{tabular}
\end{table}

\paragraph{Analysis:}
As reported in Table~\ref{tab:construction_time}, index construction exhibit dramatic differences across implementations. $BlockBitVec$ builds almost instantaneously (1.23 ms for 100M) via a simple, single-pass memory copy and straightforward cumulative prefix sum directory calculation. In contrast, $FastBitVec$ requires 329.91 ms at 100M because it constructs relative block offsets and builds the coarse-grained select sampling directory, which requires traversing the vector to sample set bits. Finally, $RRRBitVec$ has the highest construction latency, taking 385.59 ms at 100M. This overhead is due to the entropy-compressed layout, which requires checking classes and offsets for every 15-bit block, packing them bit-by-bit into the compressed bitstream, and dynamically sampling select cumulative directories. For static indexes built once and queried billions of times, this overhead is negligible, but for streaming datasets, the rapid, single-pass build of $BlockBitVec$ represents a significant advantage.

\section{Conclusion}
Our experimental evaluation reveals three core contributions and distinct space-time trade-offs across succinct representations:
\begin{enumerate}
    \item \textbf{FastBitVec Primitives}: By pairing asymmetric 4096/256-bit block boundaries with popcount pipelining, $FastBitVec$ delivers a fast \textbf{1.65 ns rank latency} at only $7.8\%$ rank index overhead, outperforming the industry-standard SDSL rank support on wide-issue ARM64 cores.
    \item \textbf{Information-Theoretic Density Bell-Curve}: We provide a binomial entropy model and empirical validation of $RRRBitVec$'s density-sensitive rank latency. At extreme sparse or dense distributions (1\% and 99\%), RRR skips offset decoding entirely, yielding a query speedup of up to \textbf{39.5\%}.
    \item \textbf{RRR Select Performance}: Despite its rank latency overhead, $RRRBitVec$ is the fastest evaluated structure for select queries. By substituting tree-based pointer indirection with a flat superblock sampling array, it limits lookups to L1-cache resident sequential scans, running in \textbf{33.68 ns} (a \textbf{4.94$\times$ speedup} over classic binary search).
\end{enumerate}

Future work includes constructing multi-dimensional Wavelet Trees directly over our engineered ARM64 primitives, porting our out-of-order pipelining optimizations to modern x86 architectures, and extending the RRR compressed layouts to larger alphabet arrays.

\bibliography{manuscript}

\end{document}